\documentclass[%
 aip,
 amsmath,amssymb,
 reprint,%
]{revtex4-1}

\usepackage{graphicx}
\usepackage{dcolumn}
\usepackage{bm}
\usepackage{physics}
\usepackage[utf8]{inputenc}
\usepackage[T1]{fontenc}
\usepackage{mathptmx}
\usepackage{etoolbox}
\usepackage[colorlinks = true,
            linkcolor = blue,
            urlcolor  = blue,
            citecolor = blue,
            anchorcolor = blue]{hyperref}

\makeatletter
\def\@email#1#2{%
 \endgroup
 \patchcmd{\titleblock@produce}
  {\frontmatter@RRAPformat}
  {\frontmatter@RRAPformat{\produce@RRAP{*#1\href{mailto:#2}{#2}}}\frontmatter@RRAPformat}
  {}{}
}%
\makeatother
\begin{document}

\preprint{AIP/123-QED}

\title{Relating the Phases of Magnetic Reconnection Growth to the Temporal Evolution of X-line Structures in a Collisionless Plasma}
\author{D.S. Payne}
 \affiliation{Institute for Research in Electronics and Applied Physics, University of Maryland, College Park, MD 20742, USA}
 \email{dspayne@umd.edu}

\author{R.B. Torbert}
\affiliation{Space Science Center, Institute for the Study of Earth, Oceans, and Space, University of New Hampshire, Durham,
NH 03824, USA}

\author{K. Germaschewski}
\affiliation{Space Science Center, Institute for the Study of Earth, Oceans, and Space, University of New Hampshire, Durham,
NH 03824, USA}

\author{T.G. Forbes}
\affiliation{Space Science Center, Institute for the Study of Earth, Oceans, and Space, University of New Hampshire, Durham,
NH 03824, USA}

\author{J.R. Shuster}
\affiliation{NASA Goddard Space Flight Center, Greenbelt, MD 20771, USA}

\date{\today}

\begin{abstract}
The efficiency of energy conversion during magnetic reconnection is related to the reconnection
rate. While the stable reconnection rate has been studied extensively, its growth between the time of reconnection onset and the peak reconnection rate has not been thoroughly discussed. We use a 2D particle-in-cell (PIC) simulation to examine how the non-ideal reconnection electric field evolves during the growth process and how it relates to changes near the x-line. We identify three phases of growth: 1) slow quasi-linear growth, 2) rapid exponential growth, and 3) tapered growth followed by negative growth after the reconnection rate peaks. Through analysis of the structural changes of the EDR, we associate the early phases with the breaking of x-line symmetry through the erosion of the pre-onset bipolar $E_z$ and the emergence of a diverging $E_x$ pattern at the neutral line in phase 1 followed by the expansion of the inflow region and the enhancement of inflow Poynting flux $S_z$ associated with the out-of-plane electric field $E_y$ in phase 2. We show how the Hall fields facilitate rapid growth in phase 2 by opening up the exhaust, relieving the electron-scale bottleneck and allowing large Poynting flux across the separatrices. We find that the rapid inflow of electromagnetic energy accumulates until the downstream electromagnetic energy density approaches the initial upstream asymptotic value in phase 3. Finally, we examine how the electron outflow and the downstream ion populations interact in phase 3 and how each species exchanges energy with the local field structures in the exhaust.

\end{abstract}

\maketitle

\section{\label{sec:level1}Introduction}

The process of magnetic reconnection consists of a topological re-configuration of magnetic fields and a transfer of electromagnetic energy to plasma energy in the form of particle acceleration and heating \citep{vasyliunas1975theoretical, sonnerup1979magnetic,yamada2007progress}.  The initiation of this process is accompanied by a localized violation of the frozen-in flux condition and a de-coupling of electron motion from the convection of the local magnetic fields.  The non-ideal electric field that violates the frozen-in flux condition is given by $\vec{E}' = \vec{E} + \vec{v}_e \cross \vec{B}$, and the energy conversion rate associated with this non-ideal electric field is given by $\vec{J}\cdot\vec{E}'$\citep{zenitani2011new}. Signatures of large positive $\vec{J}\cdot\vec{E}'$ are often used to identify the electron diffusion region (EDR) in spacecraft observations of magnetic reconnection \citep{burch2016magnetic, torbert2018electron}. While the electron-frame electric field is necessary to initiate reconnection, it does not account for all of the work done to the plasma in the reconnection process, especially outside the central EDR\citep{yamada2014conversion, yamada2015study}.  

The efficiency with which magnetic flux is reconnected and plasma is energized is described by the reconnection rate. Spacecraft observations and simulations show that the normalized rate consistently stabilizes to approximately 0.1, much faster than allowed by the Sweet-Parker reconnection model\citep{parker1973reconnection, cassak2017review}.  Recent progress has been made on the theoretical basis for this result based on pressure considerations at the reconnection site and the influence of Hall fields on the stabilized geometry of the exhaust\citep{liu2022first}. Measurements of the terms in Poynting's theorem in particle-in-cell (PIC) simulations and Magnetospheric Multiscale (MMS) data show that the EDR is typically in a steady-state energy balance, where the rate of electromagnetic flux convergence is balanced by the work rate such that the time evolution of the electromagnetic energy density is negligible \citep{genestreti2018assessing, payne2020energy}.  This steady-state balance in Poynting's theorem does not necessarily correspond with the stabilization of the reconnection rate, as it has been shown in PIC simulations that the steady-state balance occurs well before the maximum reconnection rate is reached, soon after the development of the localized out-of-plane $E'_y$ \citep{payne2020energy}.

In the present study, we do not focus on the steady-state reconnection rate, but instead on the growth of the reconnection rate during the early stages of reconnection to better understand the physical processes that accompany the transition from a quiet current sheet to stabilized fast reconnection.  Some progress has been made relating the structural evolution of the x-line and of phase space to the temporal evolution of the reconnecton rate, including the result that highly structured velocity distribution functions (VDFs) and temperature anisotropies develop in reconnection exhausts close to the occurrence of the maximum reconnection rate \citep{shuster2014highly,shuster2015spatiotemporal}.  Recent studies using MMS data \citep{hubbert2021electron,hubbert2022electron} and simulations \citep{lu2022electron} have argued that electron-only reconnection may be understood as a transitional phase between quiet current sheets and traditional electron-ion coupled reconnection in the magnetotail.  While these studies did not focus specifically on the growth of the reconnection rate, they identified multiple parameters that appeared to characterize the transition of a time-evolving electron-only current sheet, including increasing current sheet thickness, perpendicular ion temperature, parallel electron temperature, Hall field strength, and the ratio of ion to electron temperature, which grows late in the reconnection process due to the delayed heating of ions compared to electrons.  

The motivation behind the following content in this article is to better understand how the reconnection rate evolves after onset and the physical processes that control its growth. Section \ref{sec:sim} discusses the parameters of the simulation used in this study and the units associated with the variables used in the following sections. In section \ref{sec:growth}, we discuss the growth of the reconnection electric field and characterize three separate phases of the growth interval.  Section \ref{sec:structure} examines how the x-line structure changes during the first two phases of reconnection growth.  In section \ref{sec:hall}, we investigate the role of Hall fields in the rapid transport of magnetic energy and in the eventual stabilization of the reconnection rate in the last phase of reconection growth.  Section \ref{sec:species} examines the coupling between electrons and ions in the reconnection exhaust during the last phase of growth.  In sections \ref{sec:discussion} and \ref{sec:conclusions}, we discuss our results in the context of recent literature and summarize our primary conclusions.

\section{\label{sec:sim}Simulation Parameters}
This study includes a 2‐D PIC simulation with an initial Harris current sheet configuration using the plasma simulation code (PSC)\cite{germaschewski2016plasma}. The domain size is $L_x \times L_z = 80 d_i \times 20 d_i$ with 300 particles per cell (ppc) at the center of the current sheet.  The initial asymptotic magnetic field has a normalized value of $B_0 = 0.5$. There is an ion to electron mass ratio of $\frac{m_i}{m_e} = 100$ and an ion to electron temperature ratio $\frac{T_i}{T_e} = 5$. The ratio of the uniform background density to the density in the center of the current sheet is $\frac{n_b}{n_0} = 0.05$.

Here we note the units corresponding to all quantities used in the following sections.  All times presented in this paper are in units of $\Omega_{ci}^{-1}$.  Electric field contributions are in units of $\frac{c}{v_a B_0}$.  All magnetic fields are expressed in units of $B_0$.  Poynting flux is expressed as a product of $E$ and $B$, without an additional factor, so the units of Poynting flux are $\frac{c}{v_a}$.  Ion temperatures are expressed in units of $T_0$, where $T_0$ is the initial background ion temperature.  Finally, $\vec{J}\cdot\vec{E}$ and $\div \vec{S}$ are expressed in units of ${e n_0 B_0 v^2_a}$.  

\section{\label{sec:growth}Growth Phases of the Reconnection Rate}
\begin{figure}[h!]
\includegraphics[scale=0.32]{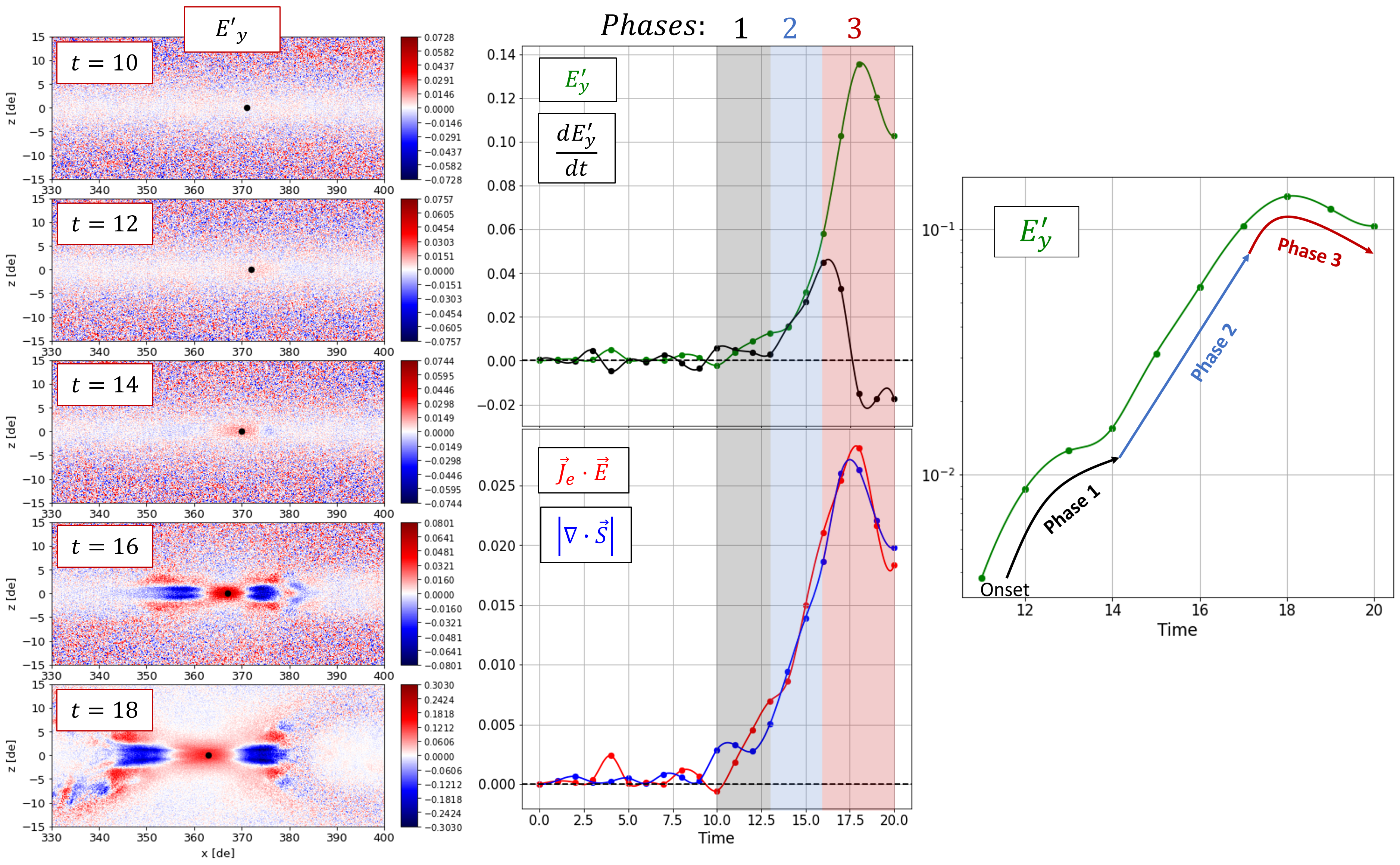}
\caption{Time evolution of parameters measured in the central EDR.  Left: EDR structure at various stages between reconnection onset and the time of the maximum reconnection rate, including a black dot indicating where quantities are measured.  Top center: The reconnection electric field and its time derivative. Bottom center: Energy transport terms including the work rate on the electrons $\vec{J}_e \cdot \vec{E}$ and the magnitude of the Poynting flux divergence $\div \vec{S}$.  Right: Semi-log plot of the reconnection electric field from $t = 11 - 20$ illustrating the three distinct phases of reconnection growth.}
\label{growth}
\end{figure}

In figure \ref{growth} we present various quantities measured at the center of the EDR as it evolves from the beginning of the simulation to a few time steps beyond the occurrence  maximum reconnection rate.  These include the reconnection electric field, which we define here as the out-of-plane contribution to the non-ideal electric field $E'_y$, and its time derivative.  We also include two quantities relevant to energy transport: the work rate associated with the energy transfer to electrons $\vec{J}_e \cdot \vec{E}$, and the divergence of Poynting flux $\div \vec{S}$.  In the EDR, $\div \vec{S}$ is negative due to the converging magnetic energy flux of the inflow regions, but we present its magnitude here in order to more easily compare to $\vec{J}_e \cdot \vec{E}$.  The magnitude of $\div \vec{S}$ in this region is therefore a measure of the rate at which Poynting flux \textit{converges} toward the x-point.

The onset of reconnection and the initial growth of $E'_y$ begins around $t = 10-11$ and reaches its maximum value at $t = 18$.  The evolution of $\frac{dE'_y}{dt}$ from onset suggests that the initial growth up to $t \approx 13$ is distinct from the growth during $t \approx 13-16$, which is distinct from the few timesteps just before and after the maximum reconnection rate, $t \approx 16-20$. Going forward, we will refer to these intervals as phases 1, 2, and 3, respectively. It should be noted that these labels are meant to be a useful way to think about different stages of the reconnection growth process, and do not apply strictly within only certain intervals (for example, $t = 13$ does not necessarily belong to only phase 1 or only phase 2, but $t = 12 - 14$ can be considered a \textit{transition} between phases 1 and 2).  

In phase 1, $\frac{dE'_y}{dt}$ is not quite constant, but is slowly decreasing with time, indicating that the reconnection rate is growing at a slightly sublinear rate.  A comparison of the energy transport terms near phase 1 suggests that the onset is preceded by an enhancement in the magnitude of $\div \vec{S}$ before the energy transfer rate $\vec{J}_e \cdot \vec{E}$ begins to grow.  The remainder of phase 1 consists of approximately linear growth in $\vec{J}_e \cdot \vec{E}$ up to $t = 13$ and approximately constant (or slightly decreasing) magnitude of $\div \vec{S}$ up to $t = 12$.  After $t = 12$, the magnitude of $\div \vec{S}$ grows at a fast rate characteristic of phase 2, but the energy transfer rate $\vec{J}_e \cdot \vec{E}$ does not transition to its phase 2 until the following timestep.  In phase 2, $\frac{dE'_y}{dt}$ grows quickly, indicating that the reconnection rate is growing at a superlinear rate.  The slope of $E'_y$ and $\frac{dE'_y}{dt}$ are similar from $t = 14-16$, suggesting that in much of phase 2, the increase in the reconnection rate is proportional to its magnitude.  This feature of exponential growth may indicate that phase 2 can be described in terms of the linear tearing mode instability.  In the transition to phase 3, $\frac{dE'_y}{dt}$ decouples from $E'_y$ as the growth in the reconnection rate begins to slow down on approach to its maximum value at $t = 18$, after which it decreases and $\frac{dE'_y}{dt}$ becomes negative.  In phase 3, both $\vec{J}_e \cdot \vec{E}$ and $\div \vec{S}$ reach their maximum magnitudes, with the maximum $\div \vec{S}$ preceding the maximum $\vec{J}_e \cdot \vec{E}$ by roughly half a time step.

On the right hand side of figure \ref{growth}, we plot $E'_y$ in semi-log order to more easily identify and illustrate the distinct phases, each of which is characterized by a different function governing the growth of the reconnection rate.  Following the onset, the reconnection rate grows at a relatively slow and approximately linear rate in phase 1.  In phase 2, the reconnection rate grows quickly.  The linear appearance of phase 2 in the semi-log plot also suggests that the growth is exponential, consistent with the linear tearing mode.  This exponential growth ceases in phase 3, as the reconnection rate approaches its maximum value and the tearing mode saturates.

To understand the physics that govern reconnection growth, it is important to relate the structural evolution of the x-line to the phases observed in the growth of the reconnection rate.  In the following section, we look at the evolution of parameters upstream and downstream of the EDR during the phases of reconnection growth.

\section{\label{sec:structure}Structural Evolution of the X-line}
\begin{figure}[h!]
\includegraphics[scale=0.45]{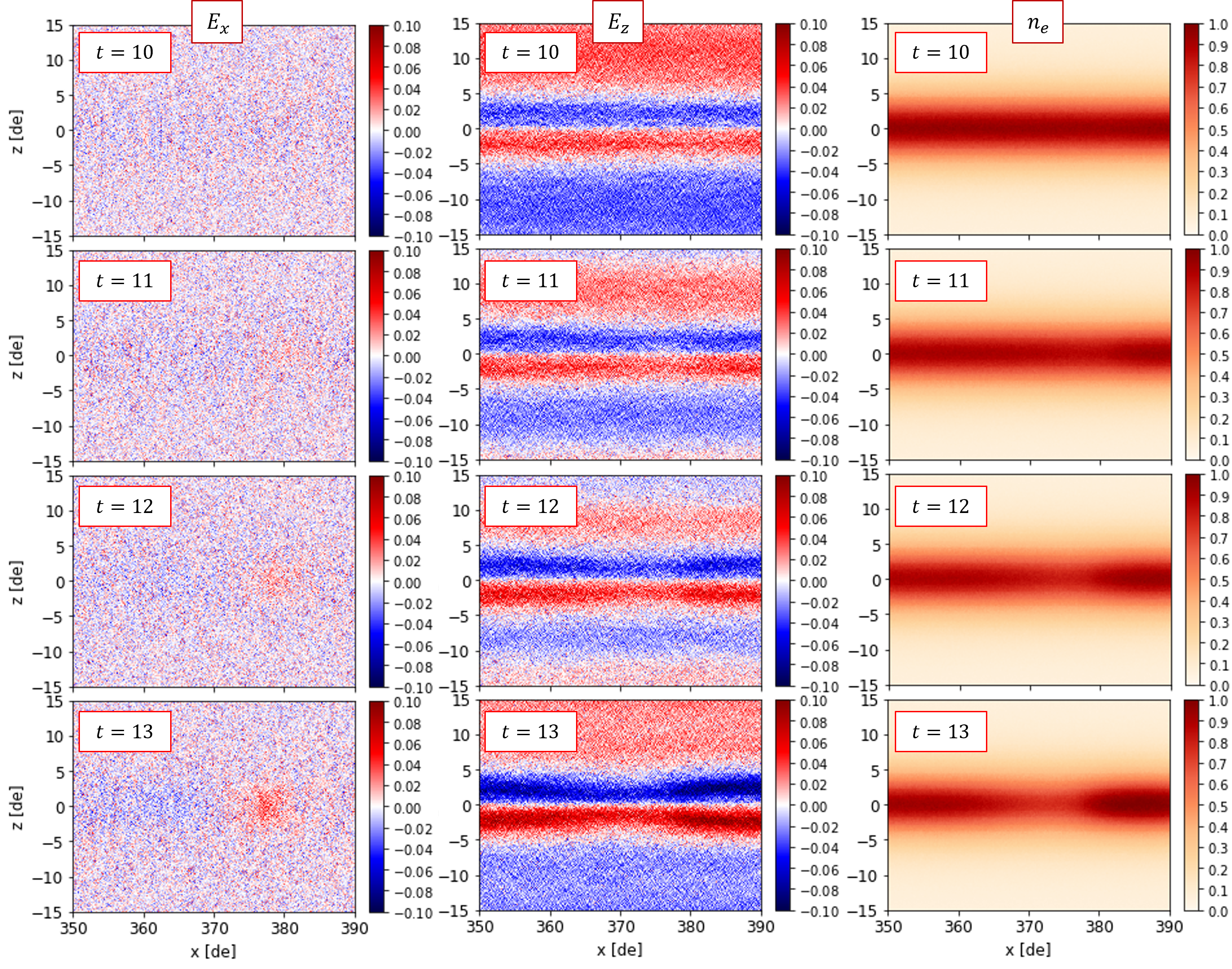}
\caption{Evolution of in-plane field components $E_x$ (left), $E_z$ (middle), and the electron number density (right) near the TCS during phase 1 of reconnection growth.}
\label{erosion}
\end{figure}

Before the onset of magnetic reconnection, there exists a uniform bipolar $E_z$ pattern along the neutral line due to the negative charging effect associated with the accumulation of electrons in the thin current sheet (TCS) \citep{pritchett1995formation,sitnov2021multiscale}.  This can be seen in the middle column of figure \ref{erosion} at $t = 10$.  The rest of figure \ref{erosion} shows how the in-plane electric fields $E_x$, $E_z$, and the electron density $n_e$ evolve during phase 1.  As phase 1 progresses, the electron density starts to diminish and the bipolar $E_z$ pattern starts to erode due to the localized loss of electron density in the TCS. The shifting of electron density away from the onset location also causes a bipolar $E_x$ pattern to emerge.  During phase 1, the symmetry along the neutral line breaks, changing the local configuration of fields and plasma. 

\begin{figure}[h!]
\includegraphics[scale=0.35]{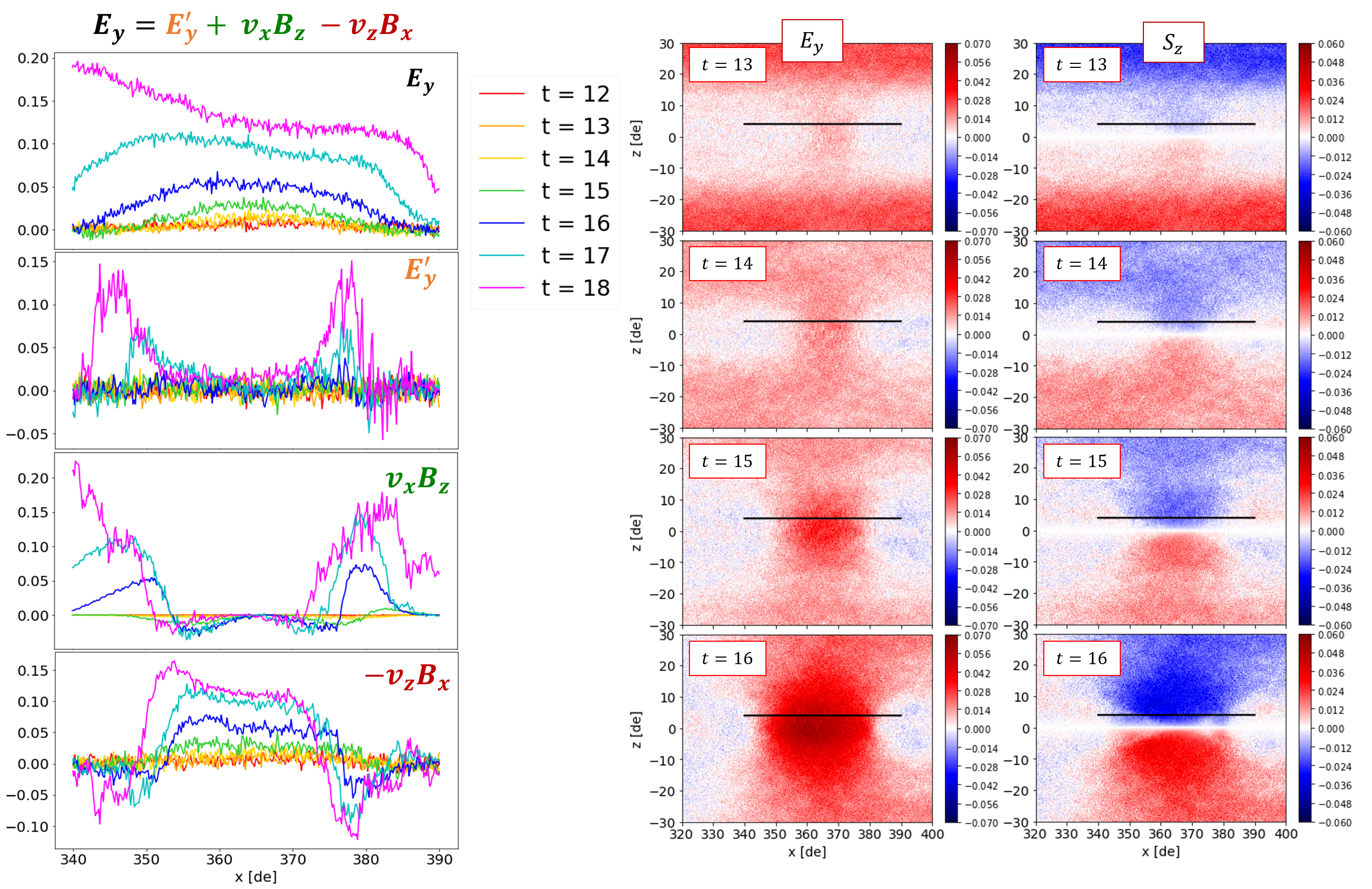}
\caption{Development of the inflow regions during phase 2 of reconnection growth.  The color plots on the right show the evolution of $E_y$ and $S_z$ during phase 2.  The panels on the left show the time evolution of multiple quantities that contribute to $E_y$, as measured along the cut shown in the color plots.}
\label{expansion}
\end{figure}

Following the onset of reconnection and the initial growth of the EDR in phase 1, a region of growing $E_y$ expands outward from the EDR toward the broader inflow and outflow regions during phase 2 as shown in figure \ref{expansion}. The expansion of the ideal $E_y$ combined with the inflow $B_x$ components produces contributions to inflow $S_z \propto E_y B_x$, allowing the electromagnetic energy transport toward the EDR to increase.  We have included in figure \ref{expansion} all potential contributions to $E_y$ along a horizontal cut shown in the inflow at $z = 4\ de$ from the end of phase 1 into phase 3.  Near the edge of the inflow, there are significant contributions from the non-ideal $E'_y$, as well as $v_x B_z$, and $-v_z B_x$. However, throughout most of the inflow region, the main contributor to the increasing $E_y$ is the increasing $-v_z B_x$ component.  The rapid enhancement of inflow $S_z$, due to enhanced out-of-plane $E_y$, is accompanied by an increasing inflow velocity pulling more electrons left over in the harris sheet toward the EDR.        

\section{\label{sec:hall}Hall Fields and Downstream Energy Transport}

\begin{figure}[h!]
\includegraphics[scale=0.35]{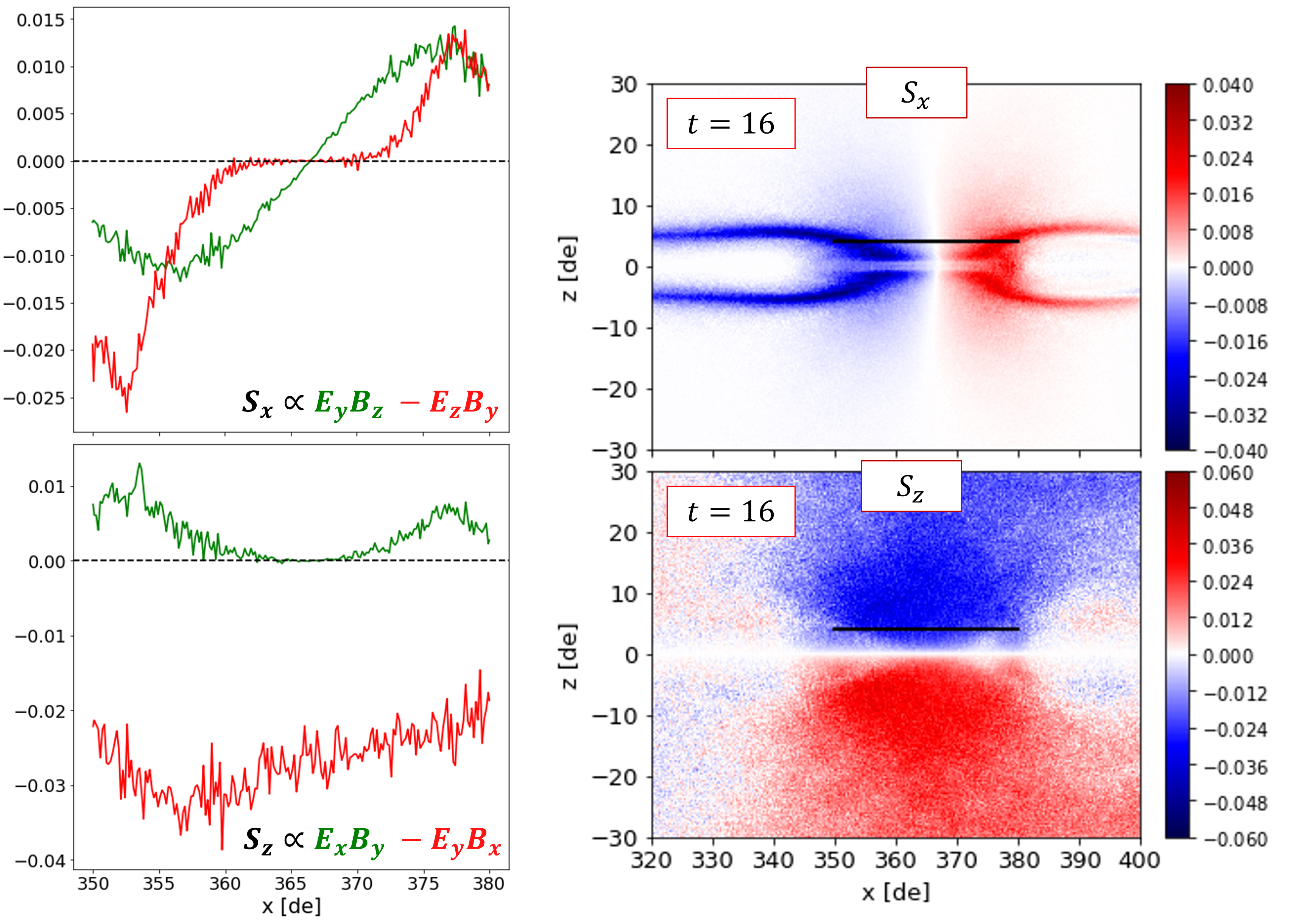}
\caption{Field contributions to the in-plane components of Poynting flux at $t = 16$.  Right: Color plots of both $S_x$ and $S_z$.  Left: Plots of the separate contributions to each component, measured along the cut shown in the color plots.}
\label{scomps}
\end{figure}

In figure \ref{scomps}, we revisit the topic of Poynting flux in the inflow. We show the separate contributions to $S_x$ and $S_z$ at $t = 16$ along a horizontal cut at $z = 4\ de$.  The cut extends across the inflow region and across the separatrices to capture both outflows. For both $S_x$ and $S_z$, the $E_y$ contribution plays a significant role, with $E_y B_z$ contributing to $S_x$ away from the EDR and $E_y B_x$ contributing to $S_z$ toward the EDR as discussed in the previous section.  These terms do not include Hall fields, and therefore reside fully within an MHD description.  However, the Hall field contributions play a non-trivial role in the production of both $S_x$ and $S_z$.  In the case of $S_x$, the contribution by $E_z B_y$ is significant across the separatrices and has the same sign as the MHD $E_y B_z$ component, thus enhancing the magnitude of $S_x$.  In the case of $S_z$, the contributions by the Hall $E_x B_y$ are also most significant across the separatrices, but have the opposite sign to the "MHD" $E_y B_x$ contribution, and therefore limit the magnitude of $S_z$.  By enhancing outflow $S_x$ and limiting inflow $S_z$, the Hall field contributions effectively change the angle and magnitude of Poynting flux across the separatrices and allow faster electromagnetic energy transport into the outflow.

\begin{figure}[h!]
\includegraphics[scale=0.3]{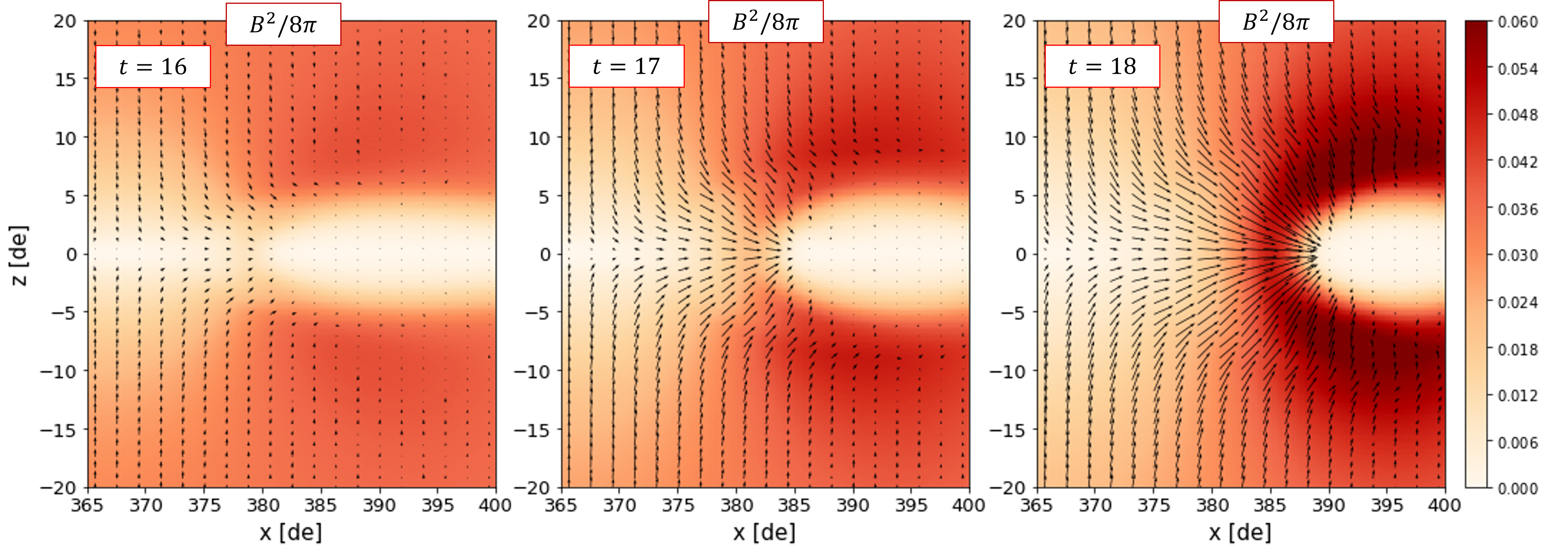}
\caption{Evolution of Poynting flux and magnetic energy density during phase 3 of reconnection growth.  The color represents the magnetic energy density and the arrows represent the in-plane Poynting flux.}
\label{hallflow}
\end{figure}

As the Hall field structures develop, they facilitate more electromagnetic energy transport by relieving the bottleneck imposed by the small scale of the EDR.  Figure \ref{hallflow} shows how this process relates to the overall transport of magnetic energy during the transition between phases 2 and 3 up to $t = 18$, when the reconnection rate reaches its maximum value.  The color in figure \ref{hallflow} represents the magnetic energy density (the electric field contribution to the electromagnetic energy density is negligible) and the arrows represent the direction and magnitude of the in-plane Poynting flux.  The arrows in all three frames are scaled to the same value, so their lengths are directly analogous to their magnitudes as they change between frames.  The Poynting flux across the separatrices continues to grow in magnitude, rapidly transporting upstream magnetic energy density into the exhaust, where it accumulates downstream of the EDR.  The upstream magnetic energy depletes and the downstream magnetic energy accumulates as the magnetic energy density in the exhaust approaches the far upstream value.  This transition to the exhaust becoming a local maximum of magnetic energy density appears to correspond with the transition to phase 3 and the time of the maximum reconnection rate.

\begin{figure}[h!]
\includegraphics[scale=0.5]{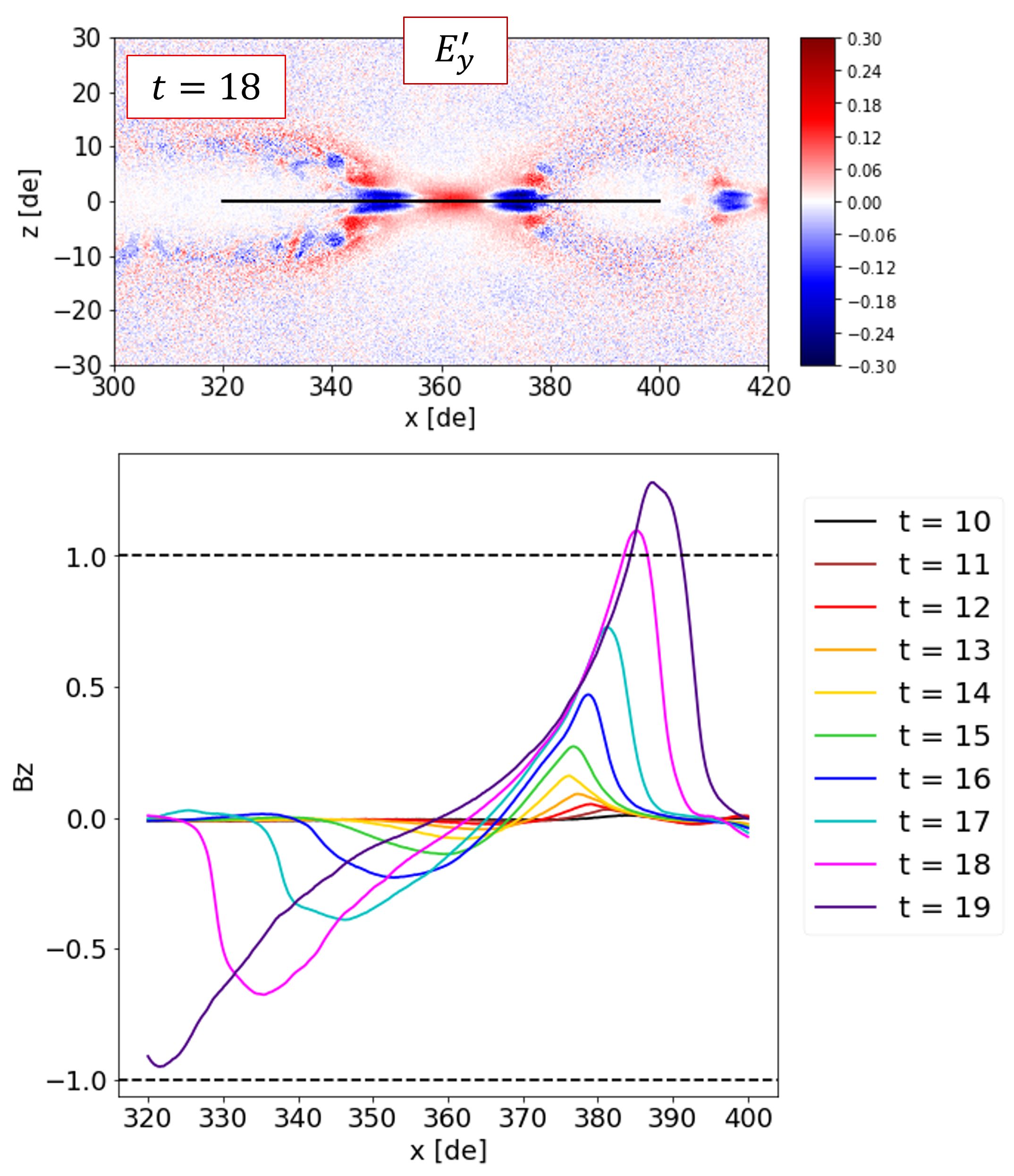}
\caption{Saturation of the downstream $B_z$.  The bottom panel shows the evolution of $B_z$, normalized to the initial maximum upstream magnetic field $B_0$,  along the cut shown in the top panel from $t = 10-19$.}
\label{saturation}
\end{figure}

Figure \ref{saturation} shows how $B_z$ along the neutral line evolves during the phases of reconnection growth.  At $t = 18$, the magnitude of the downstream $B_z$ surpasses the threshold of $B_z = B_0$ indicated by the dotted line.  While the magnitude of $B_z$ downstream continues to grow beyond $t = 18$, the growth is slower than the preceding time steps leading up to the maximum reconnection rate.  The saturation of $B_z$ occurs first on the right hand side of the outflow, whereas the saturation on the left hand side occurs two timesteps later (not shown).  This is likely due to the secondary x-line that forms downstream on the other side of the plasmoid on the right hand side, which may help accelerate the accumulation of magnetic energy in the plasmoid compared to the exhaust on the left hand side since there are two Hall field regions feeding it.  Despite this asymmetry, it appears that the occurrence of the maximum reconnection rate and the saturation of $B_z$ ($B_z \approx B_0$) along the neutral line on at least one side of the EDR are roughly correlated in time.  This is consistent with the accumulation of magnetic energy discussed previously and with the saturation of the linear tearing instability. 

\section{\label{sec:species}Electron-Ion Energy Exchange in the Exhaust}
\begin{figure}[h!]
\includegraphics[scale=0.29]{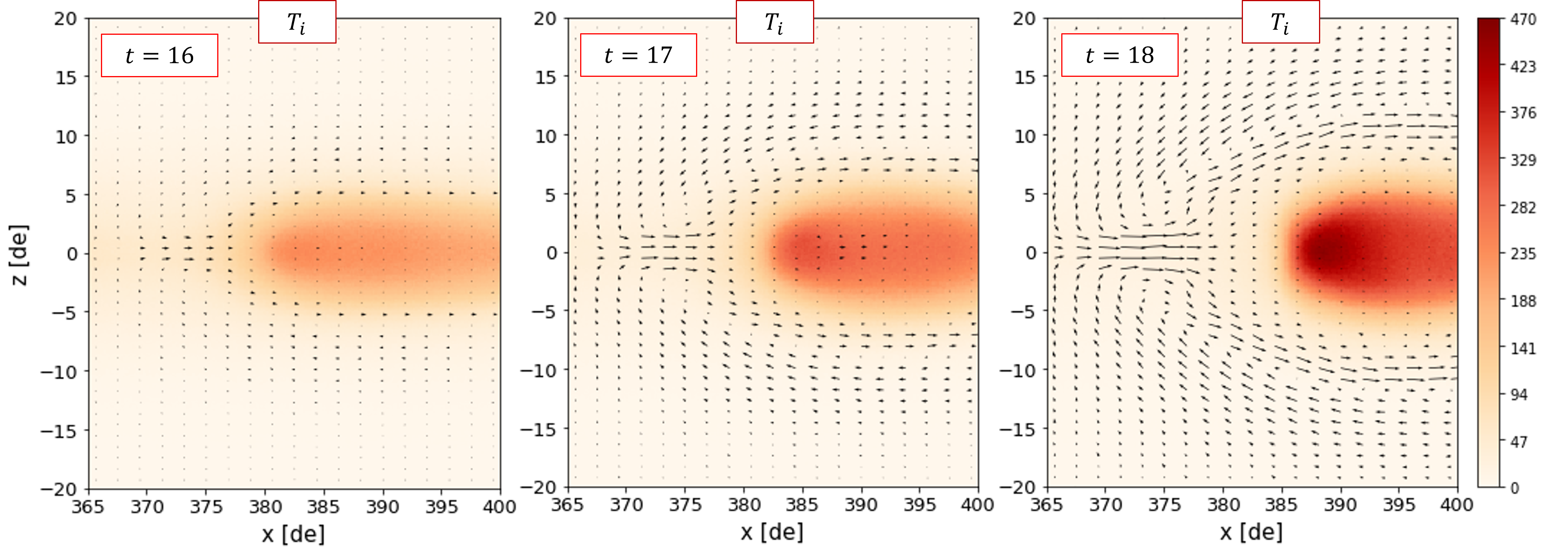}
\caption{Evolution of the electron velocity and the ion temperature during phase 3 of reconnection growth.  The color represents the ion temperature and the arrows represent the in-plane electron velocity.}
\label{species}
\end{figure}

In addition to the downstream accumulation of magnetic energy, we are also interested in how the downstream ion heating evolves along with the exhaust electron velocity.  In figure \ref{species} we show another sequence of frames from phase 3, this time showing the ion temperature overlaid with arrows indicating the in-plane electron velocity.  Just as in figure \ref{hallflow}, the arrows in figure \ref{species} are all scaled to the same value so their lengths are directly analogous to the magnitude of the in-plane electron velocity.  Both the downstream ion temperature and the electron velocity in the outflow grow in time. On the approach to the time of maximum reconnection rate ($t=18$), the region of strong ion heating starts to separate from the immediate edge of the electron jet with a region of relatively weak electron velocity in between.        

\begin{figure}[h!]
\includegraphics[scale=0.3]{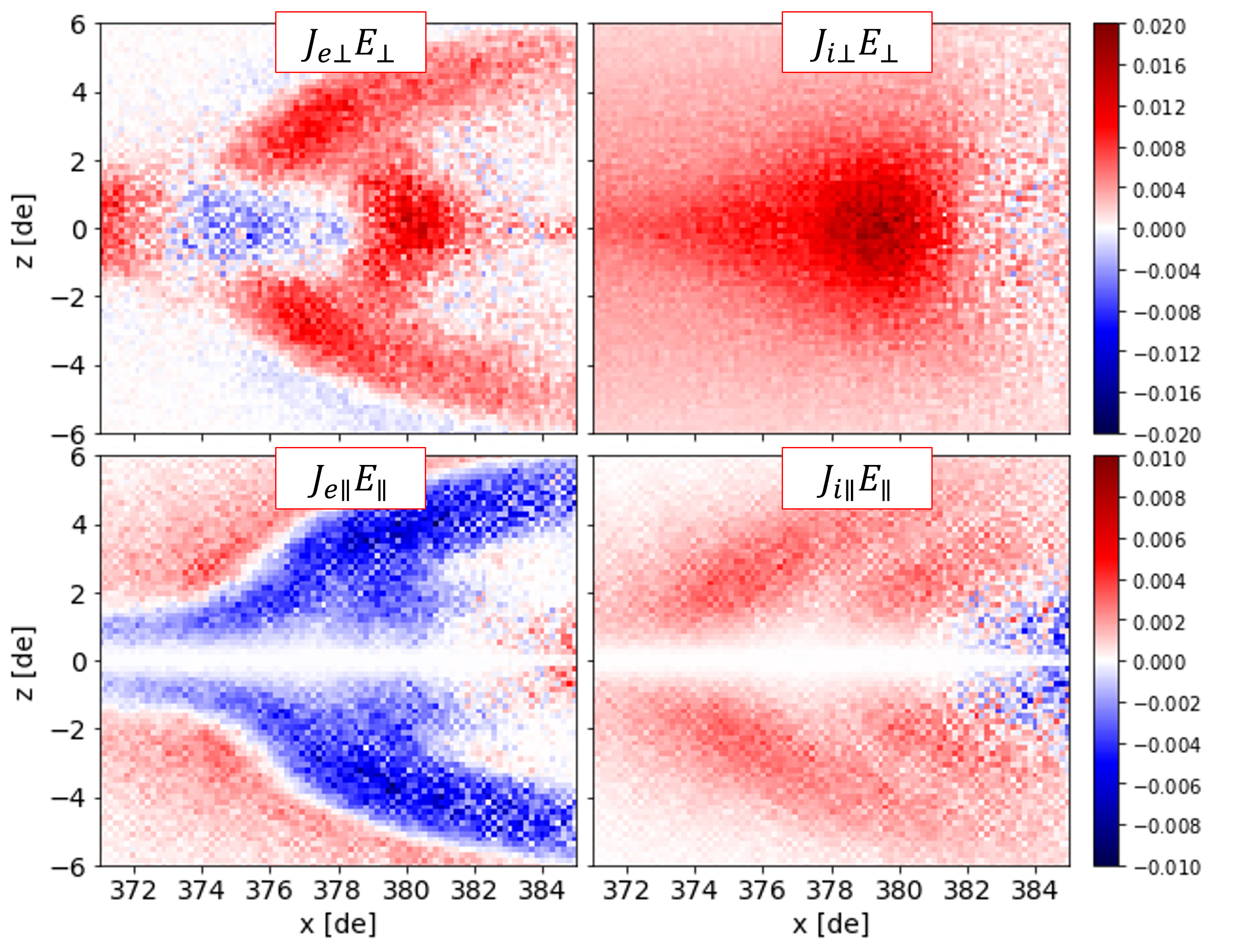}
\caption{$\vec{J}\cdot\vec{E}$ contributions broken down by species and parallel and perpendicular components at $t = 16$}
\label{parperp}
\end{figure}

To better understand energy transport and exchange in the Hall field region, we show in figure \ref{parperp} different contributions to the total $\vec{J}\cdot\vec{E}$ in the exhaust near the phase 2-3 transition at time $t = 16$.  The contributions are broken down by individual species and by components parallel or perpendicular to the local magnetic field. The electrons within the exhaust tend to lose energy parallel to the local magnetic field $J_{e\parallel} E_\parallel < 0$ and gain energy perpendicular to the local magnetic field $J_{e\perp} E_\perp > 0$ with the exception of the narrow outer EDR region, where electrons lose energy as they move across field lines $J_{e\perp} E_\perp <0$.  In contrast, the ions mostly gain energy in the exhaust, parallel and perpendicular to the local fields.     

\section{\label{sec:discussion}Discussion}
Between the onset of reconnection at the electron scale and the development of mature Petschek-type x-line structures at the ion scale, there exists a series of distinct physical processes that influence the growth of the reconnection rate, and thus the efficiency of magnetic reconnection.  Evidence of such processes can be observed in both the unique time evolution of quantities measured at the central EDR and in the structural evolution of the x-line well beyond the EDR scale.

From the reference frame of the central EDR, the onset of reconnection and the initial enhancement of $E'_y$ and $\vec{J}_e \cdot \vec{E}$ is preceded by an enhancement in the magnitude of $\div \vec{S}$, indicating a localized accumulation of electromagnetic energy density (figure \ref{growth}) based on Poynting's theorem.  This result from the same simulation was present, though not discussed, in an earlier study\cite{payne2020energy} using the same simulation which found that for most of the reconnection process the EDR exhibited an approximately time-independent balance in Poynting's theorem ($\frac{\partial u}{\partial t} \approx 0$).  The few timesteps near reconnection onset are an exception in this case where $\vec{J}_e \cdot \vec{E} \approx 0$ and the difference between the terms are not negligible compared to their magnitudes, which are relatively small at the earliest stage of reconnection growth. 

The onset of reconnection marks the beginning of what we refer to as phase 1, when the reconnection electric field in the central EDR begins to grow at a slightly sublinear rate.  In phase 1, the initial accumulation of electromagnetic energy begins to dissipate as $\vec{J}_e \cdot \vec{E}$ steadily increases (figure \ref{growth}) and some of that energy is carried away by the accelerated electrons. When $\vec{J}_e \cdot \vec{E}$ exceeds the magnitude of $\div \vec{S}$ at $t \approx 12$, the local electrons begin carrying energy away faster than the converging fields can supply it, resulting in a decrease in the local electromagnetic energy density.  Soon after this threshold, the $\div \vec{S}$ term begins to increase rapidly, followed shortly after by $\vec{J}_e \cdot \vec{E}$.  It is also during this transition to phase 2 that the growth of $E'_y$ becomes exponential, as indicated by the close agreement between $E'_y$ and its time derivative from $t \approx 14-16$ in figure \ref{growth}.  This may suggest that while the linear tearing instability plays a critical role in the growth of the reconnection rate, it is preceded by more linear processes that initiate the onset during phase 1.  After the transition into phase 2, the energy dissipation rate and the rate of field energy convergence into the region closely match and grow in lockstep, leading to a time-independent EDR even after the exponential growth and the linear tearing instability saturates.      

The structural changes near the neutral line also shed light on the processes that influence the growth of the reconnection rate in each phase.  One trivial example is the emergence of a localized nonzero $E'_y$ region (figure \ref{growth}) that accelerates electrons at the neutral line, which is how we defined the onset stage. However, the emergence of this region also influences immediate upstream and downstream parameters.  Before onset, the Harris sheet is relatively uniform along the length of the domain, but there exists a bipolar $E_z$ structure within the harris sheet that points toward the neutral line (figure \ref{erosion}), related to the accumulation of electrons into a TCS producing a negative charging effect \citep{pritchett1995formation,sitnov2021multiscale}.  We see from figure \ref{erosion} that during the slow growth of phase 1, the localized depletion of electron density at the neutral line corresponds to a localized reduction in the negative charging effect and an erosion of the bipolar $E_z$ in the immediate upstream regions. The shifting of electron density away from the onset location also produces a diverging $E_x$ pattern developing the outflow regions.  As reconnection proceeds into phase 2, the broader inflow region develops further. From figure \ref{expansion}, the $E_y$ region (not $E'_y$) grows in magnitude and expands away from the neutral line, contributing to the $S_z$ component of Poynting flux and transporting more electromagnetic energy from upstream.  The left panels of figure \ref{expansion} also highlight that the primary contribution to the growing $E_y$ and thus $S_z$ in the central inflow region is from the $v_z B_x$ term, indicating that the increased supply of electromagnetic energy is accompanied by an increase in the plasma flow toward the neutral line as expected.  In contrast, the edges of the positive $E_y$ inflow region appear to be due to mostly non-ideal effects ($E'_y$) and any residual imbalance between the in-plane flow perpendicular to the normal magnetic field ($v_x B_z > 0$)  and the in-plane flow perpendicular to the reconnecting field component ($-v_z B_x < 0$).        

\begin{figure}[h!]
\includegraphics[scale=0.36]{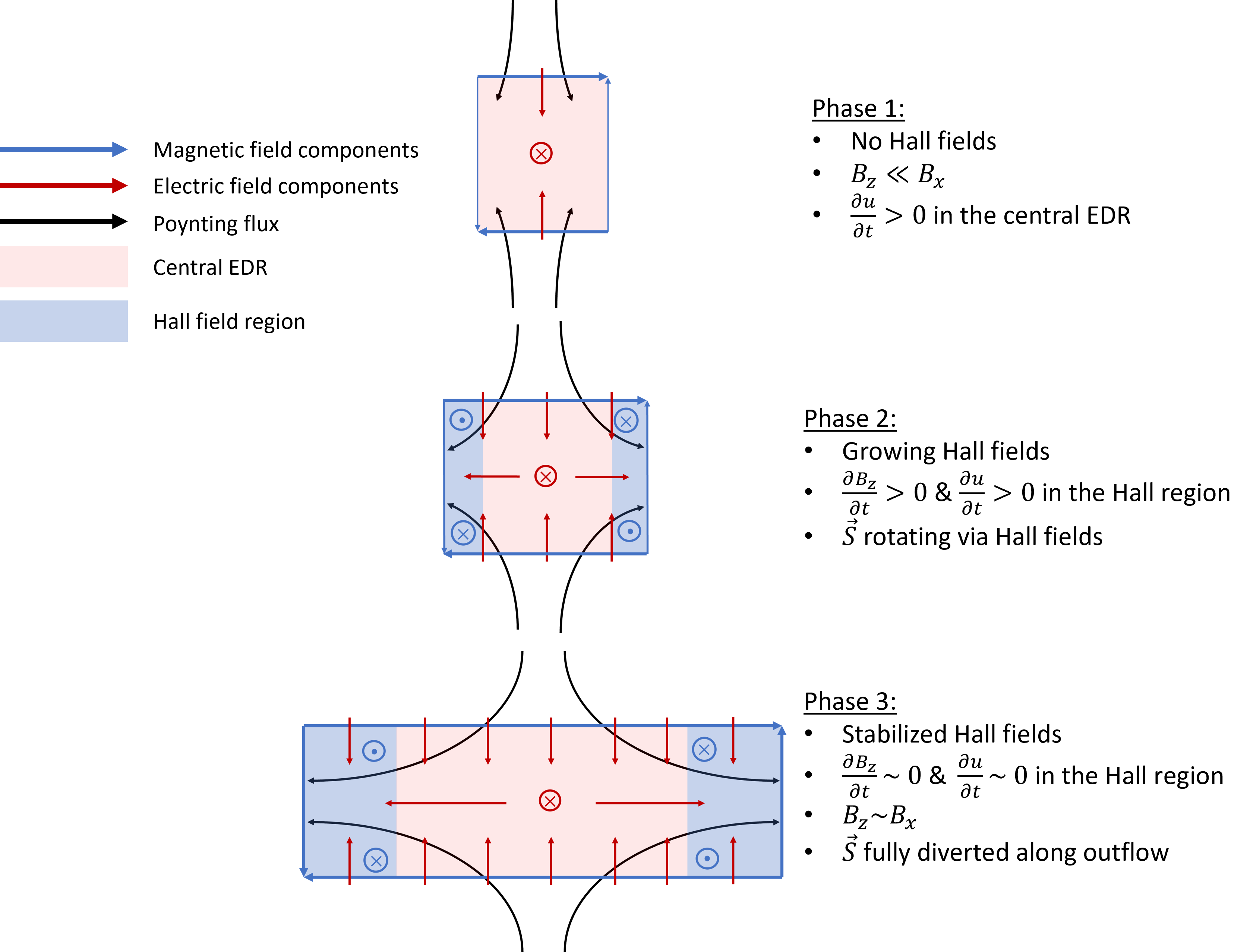}
\caption{Diagram depicting the evolution of Hall fields and energy flow across the phases of reconnection growth.}
\label{diagram}
\end{figure}

The Hall electric and magnetic fields that develop in the reconnection exhaust play an important role in the structural evolution of the x-line and the rapid increase of the reconnection rate.  In figure \ref{scomps} we showed that there are multiple processes that influence the transport of electromagnetic energy in the x-z plane.  In the absence of any in-plane electric fields or out-of-plane magnetic fields, the energy transport through the region is governed by $E_y$, $B_x$, and $B_z$.  In the case of this simulation, there is a small $\delta B_z$ component initially, so when $E_y$ begins to grow near the neutral line, the pattern of converging and diverging Poynting flux emerges from the $E_y B_x$ and $E_y B_z$ terms, respectively.  As the electron outflow and Hall currents begin to develop, the Hall fields associated with those currents change the local Poynting flux structure at the separatrices via the $E_x B_y$ term suppressing the inflow component and the $E_z B_y$ term enhancing the outflow component across the separatrices.  The net effect is that more electromagnetic energy can be transported directly across the separatrices into the exhaust because the bottleneck imposed by the small-scale EDR is relieved.  This effect of the Hall fields on the direction of Poynting flux has recently been invoked \cite{liu2022first} as a means of localizing the diffusion region around an "energy void" at the x-line and leading to Hall reconnection. In that study, they argued that within the IDR,  $\div \vec{S} \approx -\vec{J}\cdot\vec{E}_{Hall} = 0$ and $\frac{\partial u}{\partial t} \approx 0$ in the steady state\citep{liu2022first}.  This steady-state assumption is reasonable once Hall reconnection is fully developed, but here we are interested in the dynamics present as the Hall fields develop soon after the onset of reconnection.  Until the reconnection rate reaches its peak, it may be the case that $\div \vec{S} \neq -\vec{J}\cdot\vec{E}_{Hall}$ and that $\frac{\partial u}{\partial t} > 0$ in the IDR. We have seen from figure \ref{hallflow} that in the transition to phase 3 the growing Hall fields cause the magnetic energy density downstream to rapidly accumulate, eventually to energy densities larger than any present upstream as the reconnection rate reaches its peak at $t = 18$.  This threshold is also seen in the measurement of $B_z$ downstream, which grows beyond the initial far upstream $B_{x}$ value at $t = 18$ (figure \ref{saturation}).  This reasoning links the time-evolving distribution of Poynting flux induced by the Hall fields to the time-evolving magnetic energy distributions in the upstream and downstream regions.  We include a diagram in figure \ref{diagram} to illustrate the main characteristics of field evolution and energy transport during each phase of reconnection growth.  In future studies, it may be useful to examine time-evolving magnetic energy transport and accumulation not through Poynting flux, but through quantities such as magnetic flux transport (MFT) \citep{li2021identification,qi2022magnetic}, which measure the \textit{velocity} of magnetic energy transport rather than energy flux.  Such comparisons may make it easier to directly compare magnetic energy transport to plasma velocities in the outflow and examine what conditions cause the reconnection process to stabilize. 

The energy exchange between particle species in the later stages of reconnection growth has also been explored here. In figure \ref{species} we examined how the downstream ion heating and the electron flow in the exhaust co-evolve in phase 3. The buffer zone between the electron jet and the region of ion heating has relatively small electron velocity, but it corresponds to the region of large Poynting flux due to the Hall fields as shown in figure \ref{hallflow}. In a snapshot from $t = 16$, we see that near the seperatrices, electrons lose energy to $E_\parallel$ but gain energy from $E_\perp$.  This is not the case at the edge of the EDR, where $\vec{B} \approx B_z$, the local $E_\parallel$ is small, and electron jet initially loses energy to the local $E_\perp$.  More detailed explanations of the electron dynamics in the outer EDR \citep{payne2021origin,xiong2022formation} and the broader exhaust region \cite{wang2016electron} have been proposed, but will not be elaborated here.  In contrast to the electrons, the ions in figure \ref{parperp} are energized by both $E_\parallel$ and $E_\perp$ within most of the exhaust region, including the outer EDR, though the region of the largest $J_{i\perp} E_\perp$ seems to occur just beyond the outer EDR, concentrated near the same region where  $J_{e\perp} E_\perp > 0$.  These structures are likely a consequence of the different effects that Hall fields have on fast electrons leaving the EDR vs cold ions further in the exhaust.  The charge separation between the EDR and IDR boundaries creates an $E_x$ component that opposes the electron outflow, which may explain why $J_{e\perp} E_\perp < 0$ in the outer EDR and $J_{e\parallel} E_\parallel < 0$ at the separatrices.  While these fast electrons lose energy in the outflow to these electric fields, the relatively cold ions can respond to these fields and gain energy.  The Hall fields mediate energy exchange between the electron jet and the downstream ion population.   

Within the exhaust region, Hall fields play two major roles that are important to the evolution of energy transport and thus the reconnection rate.  First, they facilitate the rotation and enhancement of Poynting flux across separatrices in phase 2, opening the exhaust to allow more electromagnetic energy to accumulate and eventually saturate downstream in phase 3, leading to the stabilized fast reconnection rate of traditional reconnection.  Second, they mediate energy transfer between species by carrying energy away from the EDR and supplying some of that energy to the ions via the electric fields due to the charge separation, leading to the ion-coupling present in traditional reconnection.           

It may be useful in the future to consider how the phases of reconnection growth and the energy transport mechanisms discussed here could relate to electron-only vs traditional reconnection.  The sequence of events discussed in recent studies \citep{hubbert2021electron,hubbert2022electron,lu2022electron} appears similar to the phases of reconnection growth discussed here, especially the increasing strength of the Hall fields, which they identify as a signature of transition from electron-only reconnecting current sheets to traditional reconnection. The increase in the Hall fields during the rapid growth of the reconnection rate in phase 2 is what allows electromagnetic energy to rapidly flow into the exhaust and overcome the bottleneck imposed by the small-scale EDR.  As we also discussed here, the Hall field structures due to charge separation can also act as a means of energy exchange between energized electrons and cold ions further downstream.  This is also consistent with results\cite{hubbert2021electron,hubbert2022electron} showing that ion heating in traditional reconnection can be preceded by electron heating in electron-only reconnection. 

Another aspect worth exploring in the future is how temporal evolution of the x-line and the direction of energy flow relate to entropy, disorder, and the arrow of time.  Studies using PIC simulations \citep{liang2020kinetic,xuan2021reversibility} and MMS data \citep{argall2022theory} suggest that even in collisionless reconnection, the velocity-space kinetic entropy does increase and the reconnection process is irreversible.  Within the central EDR, electromagnetic fields do work to lower the local plasma entropy, making the local electron VDFs more ordered and less Maxwellian before those electrons become thermalized in the exhaust and increase in entropy again \citep{argall2022theory}.  The VDF structures associated with the remagnetization of the electron jet in the outer EDR \citep{payne2021origin} also appear to form near the time of the maximum reconnection rate \citep{shuster2014highly,shuster2015spatiotemporal}.  Future research may benefit from exploring how the energy-entropy exchange across the EDR relates to the flow of energy and the general trend of the reconnection process in time.

\section{\label{sec:conclusions}Conclusions}
We have examined the growth of the reconnection rate divided into three distinct phases characterized by slow quasi-linear growth, followed by rapid exponential growth, followed by a tapering and eventual reduction of the reconnection rate following its peak.  Based on the structural changes early after onset, we found that phase 1 is associated with the breaking of the neutral line symmetry, the reconfiguration of the $E_z$ pattern due to the negative charging of the TCS, and the emergence of an $E_x$ pattern diverging from the onset region.  Following phase 1, the inflow region grew in both spatial extent and in magnitude of inflow Poynting flux $S_z$ and $E_y$, accompanied by an increase in the inflow velocity.  By comparing different field contributions to the $S_x$ and $S_z$ components across the inflow region, we showed how the Hall electric and magnetic fields play a role in the rapid transport of magnetic energy downstream and thus the rapid growth of the reconnection rate.  While the emergence of the Hall fields initiated phase 2, the accumulation and saturation of the energy they transported downstream of the EDR marked the transition to phase 3, limiting further growth of the reconnection rate.  Finally, we examined the interaction between different particle species via energy exchange with parallel and perpendicular electric fields in the Hall region, highlighting how energy exchange between species plays a role in the development of the Hall region of the reconnection exhaust.

\begin{acknowledgments}
This work also depends on those who maintained and operated Trillian at UNH. Funding for this work comes from NASA via Grants NNG04EB9C and NNX13AK31G as well as NSF funding via NSF-1460190. J.R. Shuster was supported by NASA grants 80NSSC21K0732 and 80NSSC21K1482. I would also like to thank Matthew Argall for helpful discussion when I started this work.
\end{acknowledgments}

\section*{Data Availability Statement}
The simulation used in this study was run on the CRAY supercomputer Trillian at the University of New Hampshire in Durham, NH. The simulation data are archived in a Zenodo repository (DOI https://doi.org/10.5281/zenodo.4022436).

\section*{References}

\nocite{*}
\bibliography{RxEv}

\end{document}